\documentclass[conference]{IEEEtran}
\usepackage[utf8]{inputenc}
\usepackage{graphicx}
\usepackage{epstopdf}
\usepackage{amsfonts}
\usepackage{amsmath}
\usepackage{color}
\usepackage{cite}
\usepackage{mathtools}

\usepackage{stmaryrd}
\usepackage{algpseudocode}% http://ctan.org/pkg/algorithmicx
\usepackage[linesnumbered,ruled,vlined]{algorithm2e}
\usepackage{url}
\usepackage[thinlines]{easytable}
\usepackage{makecell}
\usepackage{subfigure}

\usepackage[linesnumbered,ruled]{algorithm2e}

\newcommand*\diag {\mathop{}\!\mathrm{diag}}  
\newcommand*\de {\mathop{}\!\mathrm{de2bi}}

\DeclareMathOperator*{\argmaxA}{arg\,max} % Jan Hlavacek
\DeclareMathOperator*{\argminA}{arg\,min} % Jan Hlavacek
\ifCLASSINFOpdf

\else
\fi
\begin{document}
\title{ Novel  Codebook-based MC-CDMA with Compressive Sensing Multiuser Detection for Sporadic mMTC}
\author{\IEEEauthorblockN{Mehmood Alam, Qi Zhang}
\IEEEauthorblockA{Department of Engineering \\
	 Aarhus University, Denmark
\\Email:\{mehmood.alam, qz\}@eng.au.dk
}}
\maketitle
\begin{abstract}
Massive machine type communication (mMTC) is one of the basic components of the future fifth generation (5G) wireless communication system. In mMTC, the information processing at the sensor nodes is required to be simple, low power consuming and spectral efficient. In order to increase the spectral efficiency at the cost of minimum performance loss and to simplify the information processing at the sensor node, in this paper a codebook based spreading technique is proposed. In the proposed scheme, first, the modulation and spreading are combined into a direct symbol-to-sequence spreader, which directly maps the input bits into a codeword from the user specific codebook. Secondly, utilizing the sporadic node activity of the mMTC a compressive sensing based algorithm is designed for multiuser activity and data detection. Exploiting the multidimensional structure of the codebook, the required signal to noise ratio (SNR) to increase the modulation order, is reduced as compared to the conventional scheme. Besides reducing the power consumption, it is shown that for modulation order of eight, the proposed scheme achieves  a gain of 1 dB over the conventional scheme at bit error rate (BER) of $10^{-5}$. However, more sophisticated decoding algorithm is required to achieve the performance gain at a lower modulation order.\par
 \end{abstract}
\IEEEpeerreviewmaketitle
\section{Introduction}
It is estimated that the number of smart devices will rise to 50 billion \cite{evans2011internet} by 2020 and even up to 500 billion by 2024 \cite{500}.
 This substantial increase in the number of devices imposes enormous challenges  to meet the performance requirements such as higher spectral efficiency, massive connectivity,  low power consumption, ultra-high reliability and ultra-low latency of the fifth generation (5G) communication system \cite{zhang2015mission} \cite{wp}.
 The current orthogonal multiple access (OMA) schemes allocate orthogonal resources in frequency, time or code domain to separate the users. Therefore, the capability of the OMA schemes to accommodate the massive number of devices in the network is limited by the total number of orthogonal resources.  In order to meet the requirement of massive connectivity, the focus of current research is towards non-orthogonal multiple access (NOMA) schemes. In NOMA schemes, non-orthogonal resources are allocated to the sensor nodes, which allow more nodes to transmit at the same time. Note that node and user are used interchangeably in this paper.  \par 
 NOMA schemes can be divided into two main groups: power domain and code domain. In power domain NOMA \cite{pnoma}, the users are allocated with different power levels to transmit their data. At the receiver, successive interference cancellation (SIC) \cite{sic} is employed, which uses the different power levels to separate the users. In code domain NOMA, the users are assigned with non-orthogonal codes to transmit their data. At the receiver, more sophisticated multiuser detection (MUD) techniques such as message passing algorithm (MPA) \cite{mpa} and SIC are used to separate the users. The prominent code domain NOMA schemes are sparse code multiple access (SCMA) \cite{scma, MC, 7925767}, pattern division multiple access (PDMA) \cite{pdma}, multi-user shared access (MUSA) \cite{musa} and so on. In SCMA, the information bits after channel coding are directly mapped to user specific sparse codewords and MPA is used at the receiver for MUD. The PDMA is also codebook based NOMA that considers each user’s channel state and accordingly allocates a different number of non-zero elements to the codeword. The MUSA is a sequence based NOMA scheme that assigns low correlated spreading sequences to the nodes and uses SIC for MUD. 
  All these NOMA schemes allow more users to connect to the network by allocating non-orthogonal resources. However, in all of these schemes, it is assumed that all users are active at the same time, whereas the mMTC is highly sporadic, i.e., only a small fraction of the users is active at a time. Besides the sporadic nature, the mMTC is also characterized by low data rate transmissions, which necessitate a grant free medium access control mechanism to reduce the control signaling overhead.\par
   Compressive sensing based multiuser detection (CSMUD) \cite{csp1, 6125356, 6240301, DBLP, ECSMUD} is a recently proposed scheme to enable grant free non-orthogonal code division multiple access (CDMA) for sporadic mMTC.  Non-orthogonal spreading sequences are assigned to the users which facilitate multiple users  to share the same physical resource. At the receiver, CSMUD uses compressive sensing algorithms to jointly detect the activity and data. A comprehensive study of the  CSMUD is given in our review article \cite{csmudservy}. \par
In the typical non-orthogonal CDMA, the data after channel encoding is modulated using modulation techniques, e.g., phase shift keying (PSK). The BER performance depends on the minimum Euclidean distance of the PSK constellation.  To enhance the spectral efficiency by increasing the modulation order of the PSK modulator, the minimum Euclidean distance of the PSK constellation decreases, which results in increasing the  BER.  Furthermore, the input bits to the PSK modulator pass through a series of components such as bit splitter, level shifters, reference oscillator, product modulators, band pass filters and linear summer\cite{psk1}. The information processing in these components contributes in increasing the latency and power consumption  of the hardware.\par 
To overcome the above limitations, in this paper we first propose a codebook based direct spreading scheme for the non-orthogonal multicarrier-CDMA (MC-CDMA) \cite{mccdma,MC} at the transmitter. Secondly, a modified group matching pursuit (GMP) algorithm is proposed for joint multiuser activity and data detection at the receiver. The PSK modulator and the CDMA spreader are combined into a symbol-to-sequence spreader, which maps the input set of bits into a codeword (spreading sequence) in the user specific codebook. The codebook of each user consists of the circularly shifted versions of the user specific spreading sequence. Besides simplifying the information processing at the sensor node, the codebook based spreading reduces the performance degradation for increasing the modulation order. Simulation results shows that  the proposed scheme achieves  a gain of 1 dB over the conventional scheme for $M=8$ at BER of $10^{-5}$. Furthermore, elimination of modulator simplifies the information processing and increases the battery life of the sensor node. Note that modulation order for the direct spreading scheme means that there are $M$ number of possible combinations of input bits. \par  
 The paper is organized as follows. In Section \ref{CSbasics}, the basic compressive sensing theory is given, Section \ref{sysmod} presents the general system model for mMTC. In Section \ref{proposed}, the proposed method of direct symbol-to-sequence spreading, design of the codebooks and the joint multiuser activity and data detection are described. In Section \ref{pa}, the simulation parameters are given and performance of the proposed scheme is analyzed in terms of BER. Finally Section \ref{con} concludes the paper and suggests future work.\par
 \textit{Notations:} In this paper, all boldface uppercase letters represent matrices such as $\mathbf{S}$, while all lowercase boldface letters represent vectors such as $\mathbf{s}$, $\mathbf{x}$. The set of binary, integers and complex numbers are represented by $\mathbb{B}$, $\mathbb{Z}$ and $\mathbb{C}$, respectively.  Italic letters such as $k$, $x$ represent variables. Uppercase letters such as $K$ and Greek letters such as $\gamma$ represent a constant value and $\Gamma^ {-1}(.)$ is an inverse indexing operator.   
 \section{Compressive sensing Basics} \label{CSbasics}
 Compressive sensing (CS) is a signal processing technique which samples a sparse signal at a rate much less than the Nyquist rate \cite{cs}. A signal  $\mathbf{v}\in\mathbb{C}^{K\times1}$ is said to be $K_a$-sparse if it has only $K_a$ non-zero elements. The CS process produces measurement $\mathbf{y} \in \mathbb{C}^{N\times 1}$ by a measurement matrix  $\mathbf{\Psi}\in \mathbb{C}^{N\times K}$, $K_a<N<K$,
 \begin{equation}
 \mathbf{y}=\mathbf{\Psi}\mathbf{v}.
 \label{cseq}
 \end{equation}\par 
 Equation (\ref{cseq}) is an underdetermined system of equations, therefore, the vector $\mathbf{v}$ is recovered  by convex minimization as follows
 \begin{equation}
 \hat{\mathbf{v}}= \argminA_{\mathbf{v} \in \mathbb{C}^{N}} {\Vert \mathbf{v} \Vert}_1  \:  \text{subject to } \:  \mathbf{y} =\mathbf{\Psi} \mathbf{v},
 \end{equation}
 where ${\Vert \mathbf{v} \Vert}_1 $ represents the $l_1$ norm of $\mathbf{v}$.  $l_1$ minimization efficiently recovers the signal, however, it has a complexity of cubic order. To reduce the complexity of the sparse signal recovery, greedy algorithms such as orthogonal matching pursuit (OMP) \cite{omp1} and orthogonal least square (OLS) are used for signal recovery \cite{cs}. Greedy algorithms iteratively obtain the support for the received signal. In OMP the support of $\mathbf{v}$ is obtained by selecting the column of $\mathbf{\Psi}$ which has maximum correlation with the residual, whereas in OLS the column with least square distance from the residual is selected. The residual is initialized as $\mathbf{y}$ and is updated in each iteration of the greedy algorithms. Once the support is obtained, the corresponding value is estimated by using least square estimation. For the successful recovery of signal, the measurement matrix, $\mathbf{\Psi}$, needs to satisfy the restricted isometry property (RIP) \cite{candes2005decoding}.   Generally verifying the RIP for a sensing matrix is NP-hard, however, a Guassian matrix is known to satisfy RIP  with high probability whenever the number of rows scales linearly with the sparsity  of the signal and logarithmically with the length of the signal \cite{bah2010improved, candes2006compressive}.\par  
\section{System Model} \label{sysmod}
\begin{figure}[h]
	\centering
	\includegraphics[scale=.7]{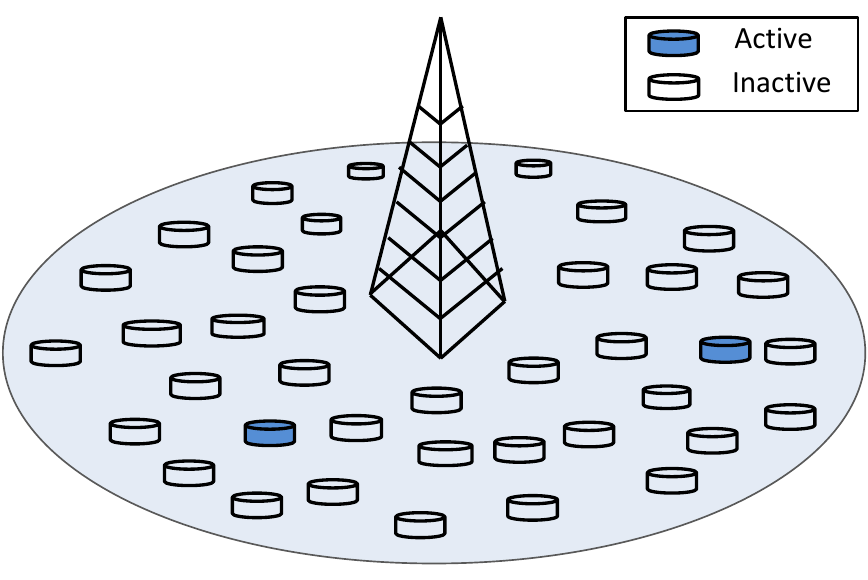}
	\caption{ 	\label{sc} Massive Machine type communication: $K$ nodes connected to the base station}
\end{figure}

Consider a typical uplink mMTC scenario where $K$ sensor nodes are in the proximity of the aggregation node as shown in Figure \ref{sc}. 
The data transmission of nodes is sporadic, i.e., very few of the nodes are active at a given instant of time. Furthermore, we assume that when a node is active, it transmits $L$ consecutive bits. Under these assumptions, the mMTC traffic is modeled as a Bernoulli distribution with activity probability, $p_a\ll1$, and inactive probability $1-p_a$. This sporadic nature of mMTC allows us to model the multiuser detection as a compressive sensing problem.\par

For mMTC in the non-orthogonal MC-CDMA system, the column vector $\mathbf{v}$ in Equation (\ref{cseq})  is the multiuser signal which represents the data symbols of $K$ users at time instant, $t$, $K_a=p_aK$ and the sensing matrix $\mathbf{\Psi}$  consists of the combination of the spreading and channel matrices. 
\section{Direct symbol-to-sequence spreading} \label{proposed}
\begin{figure}[h]
	\centering
	\includegraphics[scale=.52]{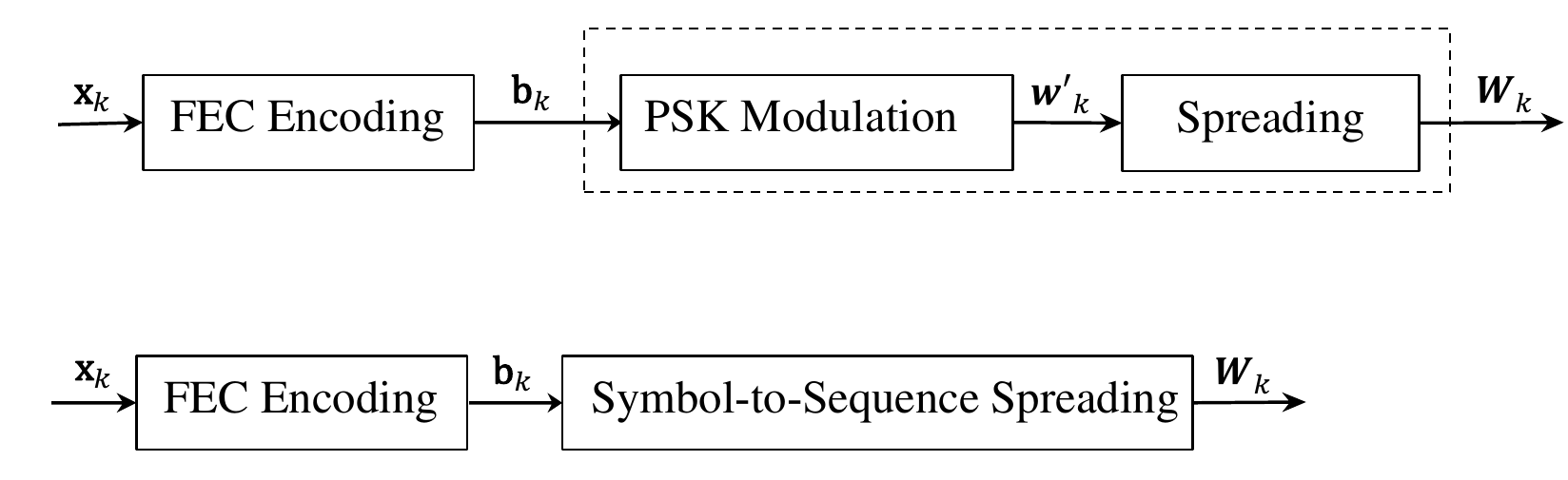}
	\caption{Merging of PSK modulation and spreading into symbol-to-sequence spreading}
	\label{enc}
\end{figure}

The signal generation at a sensor node $k$ starts with encoding a stream of $L$ bits, $\mathbf{x}_k \in \mathbb{B}^{1 \times L}$, into a stream of $L_c$ coded bits $\mathbf{b}_{_k} \in \mathbb{B}^{1 \times L_c}$. In conventional non-orthogonal MC-CDMA, the channel coded frame, $\mathbf{b}_k$,  is first modulated and then  spread over the spreading sequence to obtain the spread data matrix, $\mathbf{W}_k\in\mathbb{C}^{N\times L_d}$, where $L_d=\frac{L_c}{\log_2 (M)}$,  $N$ is the length of the spreading sequence and $M$ is the modulation order. In the proposed scheme the PSK modulation and the spreading are merged together and the set of incoming bits are directly mapped to a spreading sequence based on the user specific codebook as shown in Figure \ref{enc}.\par 
Let $\mathbf{d}_k \in \mathbb{Z}^{1\times L_d}$ represent the $\log_2 (M)$ bits decimal counterpart of  the channel coded frame, $\mathbf{b}_{_k}$, and $\mathbf{p}_{_k}$ represents the shifting pattern of node $k$.  The direct symbol-to-sequence spreader is defined as 
\begin{equation}
\nonumber
f:\mathbf{s}_{_k} \longrightarrow   \mathbf{s}_{_k}^{( \Gamma^{-1}(\mathbf{p}_{_{k,d_{_{k,i}}}} ) )} , 
\end{equation}
where  $ \mathbf{s}_{_k}^{( \Gamma^{-1}(\mathbf{p}_{_{k,d_{_{k,i}}}} ) )}$ is the sequence, $\mathbf{s}_{_k}$, shifted by $  \Gamma^{-1}(\mathbf{p}_{_{k,d_{_{k,i}}}} )$, $ 1 \le i \le L_d $. The inverse indexing operator, $  \Gamma^{-1}(\mathbf{p}_{_{k,d_{_{k,i}}}} )$, gives the value at the  $d_{_{k,i}}$-th index of the shifting pattern,  $\mathbf{p}_{_k}$.  For example, the shifting pattern for node 3 is $\mathbf{p}_{_3}=[0,3, 15, 6]$ and the data symbol at index, $i=2$, is $d_{_{3,2}}=4$, then  $  \Gamma^{-1} {(\mathbf {p}_{_{_3,d_{_{3,2}}}} )}=6$. The direct symbol-to-sequence encoder will shift the sequence $\mathbf{s}_{_3}$  by 6.\par
 Unlike the conventional non-orthogonal MC-CDMA where low correlated spreading sequences are assigned to the users, here each user is assigned with a unique codebook that is known at the receiver. The codebook consists of a group of spreading sequences, which are the shifted versions of the base sequence, i.e., the spreading sequence assigned to a user. The base sequence is shifted according to a user specific shifting pattern, which is discussed in the next subsection. The input set of bits is mapped to a sequence in the codebook according to the combination of the bits. For example for the case of two bits set, (00) can be  mapped to the original sequence, (01) to the sequence circularly shifted by one, (11) to the sequence circularly shifted by two and so on as shown in Figure \ref{exp1}.
 \begin{figure}[h]
 	\centering
 	\includegraphics[scale=.8]{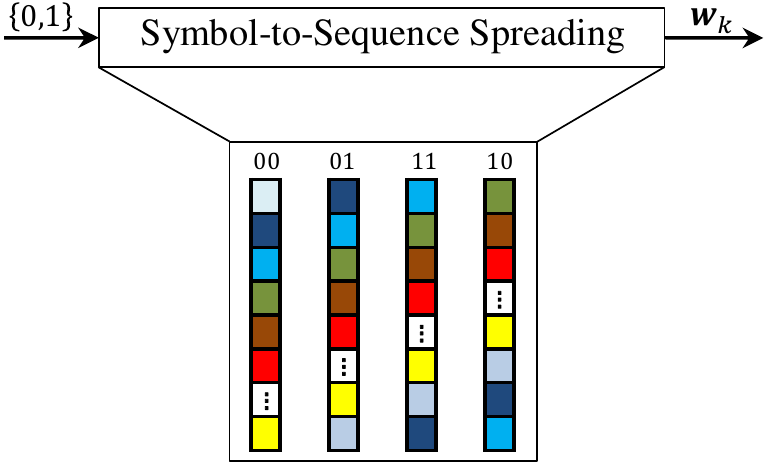}
 	\caption{An example of symbol-to-sequence spreading: two bits mapped to a sequence}
 	\label{cbb1}
 \end{figure}
 The selected spreading sequences from all active nodes are then multiplexed using orthogonal frequency division multiplexing (OFDM). \par  
 In the conventional scheme, the data after despreading is fed into the PSK demodulator. The performance of the PSK demodulator is limited by the minimum Euclidean distance ($d_{min}$) between the two dimensional constellation points. For MPSK,  $d_{min}$ is given as  \cite{pskk}
\begin{equation}
d_{min}=2\sqrt{E_s}\sin(\frac{\pi}{M}),
\end{equation}
 where $E_s$ is symbol energy and $M$ is the modulation order. As the modulation order increases, the minimum distance between the constellation points decreases which increases the error probability. For the direct spreading, the data is directly detected based on a multidimensional codebook, which can be regarded as a multidimensional constellation.  In the proposed scheme due to the multidimensional codebook the reduction in minimum Euclidean distance is much less as compared to that of  a two dimensional constellation. Therefore, for increasing the modulation order, the BER of the proposed scheme does not increase significantly, which results in a gain in the spectrum efficiency as we scale up the modulation order. The error probability for the proposed scheme is determined by the minimum correlation between the spreading sequences of the codebooks and by the multiple access interference (MAI). However, due to the sporadic transmission in mMTC the number of active nodes is far less than the total nodes, which restricts the MAI. \par
Eliminating the PSK modulator simplifies the information processing as the spreading is now directly done by mapping the input symbol into a codeword, which consequently reduces the power consumption of the node. Due to different implementation of the PSK modulator in different sensor nodes, the reduction in power consumption varies, however, it is more significant in short-range mmMTC where the transmission power is low.  The reduction in power consumption increases the battery life which can be simply estimated by the battery capacity in mAh and the current consumed in mA. For example, the CC2420\cite{datasheet20062} consumes 17.4 mA current at a transmission power of 0 dBm. Removing the PSK modulator, which consumes 5 mW power \cite{sasilatha2009comparative}, the battery life of CC2420 transceiver can be increased by approximately 8\%. \par
\subsection{Sensing matrix and shifting pattern design}
 The sensing matrix  $\mathbf{S}=\begin{bmatrix} \mathbf{s}_{_{1}}\!\!\!&\mathbf{s}_{_{2}}\ldots \mathbf{s}_{_{k}}\ldots  \mathbf{s}_{_{K}}\end{bmatrix}$, is obtained  by selecting random sequences from the unit circle such as $s_i(\mu)\sim \exp(2\pi\mu)$ with $\mu$ being a uniform distribution on the interval [0,1]. For each sequence,  $\mathbf{s}_k$,  the shifting pattern,  $\mathbf{p}_{_k}$,  is obtained such that when $\mathbf{s}_k$ is shifted according to the pattern, the   shifted versions of the sequence have minimum mutual correlations.  $\mathbf{p}_{_k}$ for the spreading sequence, $\mathbf{s}_k$,  is obtained by setting ${p}_{_{k,1}}=0$ and selecting the other entries as 
\begin{equation}
	{p}_{_{k,m}}= \argminA_{{j,\: 1\leq j \leq {N-1}}} {\left(  \sum_{i=1}^{m-1  }  \langle\mathbf{s}_k^{({p}_{_{k,i}})}, \mathbf{s}_k^{(j)} \rangle   \right)},  2\leq m\leq M , 
\end{equation}
where  ${p}_{_{k,m}}$ is the value at the $m$-th index of the shifting pattern of user $k$. $\mathbf{s}_k^{(j)}$ represents the $j$ circularly shifted version of  $\mathbf{s}_k$ and $\langle\mathbf{s}_k^{({p}_{_{k,i}})}, \mathbf{s}_k^{(j)} \rangle$ is the correlation between $\mathbf{s}_k^{({p}_{_{k,i}})}$ and $ \mathbf{s}_k^{(j)}  $. \par 
In \cite{zhang2011sparse, wang2016exact } it is shown that the OMP recovers the signal if $M \ge O (K_a \log K)$. In the proposed scheme, the number of active nodes, $K_a$, remains the same as that in conventional, however, the number of spreading sequences, $K$, is increased by the size of the codebook, i.e., by $M$. Therefore, for the proposed scheme, the condition on the number of measurements for signal recovery is given as 
\begin{equation}
M \ge O (K_a \log MK).
\end{equation}
\subsection{Multiuser detection}
At the receiver, the received signal, $\mathbf{Y}\in \mathbb{C}^{N\times L_d}$,  after being affected by  fading channel and additive white Guassian noise (AWGN) can be represented in frequency domain as
\begin{equation}
\mathbf {Y}  =  \sum_{k=1}^{K} \diag(\mathbf{h}_k) (\mathbf{s}_k \otimes  \Gamma^{-1} ( \mathbf{p}_ { _ {k,\mathbf{d}_k} } ) )  + \mathbf{N},
\label{sm}
\end{equation} 
where $\mathbf{s}_k$ is the spreading sequence of node $k$ and $\mathbf{N}$ is the AWGN noise matrix.  $\diag(\mathbf{h}_k)$ is the channel matrix of user $k$  which has channel coefficients at its diagonal, where $\mathbf{h}_k$ is a column of the matrix  $\mathbf{H} \in \mathbb{C}^{N \times K}$. Vector $\mathbf{d}_k$ is the $k$-th row of the matrix $\mathbf{D} \in \mathbb{Z}^{K\times L_d}$. Each row of  matrix  $\mathbf{D}$ is the data frame of a user and each column consists of the data symbols of $K$ users at time $t$. The inactive users are represented by all zero rows of $\mathbf{D}$. The operator $\otimes$ is defined as $\mathbf{a} \otimes \mathbf{c} = [ \mathbf{a}^{(c_1)}\:\mathbf{a}^{(c_2)} \hdots \mathbf{a}^{(c_t)} \hdots \mathbf{a}^{(c_{_T})}]$, where   $\mathbf{a}^{(c_t)}$ is the $c_t$ times circularly shifted version of $\mathbf{a}$, $c_t$ is the value at the  $t$-th index of $\mathbf{c}$  and $T$ is the length of vector $\mathbf{c}$. \par 
The support is obtained iteratively using a modified GMP algorithm. The active users are detected iteratively based on the mean  of the maximum correlation between the spreading sequences of the user's codebooks and the received OFDM symbols. After the active users are detected, least square  estimation is used to detect the data. The algorithm is given in Algorithm  \ref{alg:the_alg}.
\begin{algorithm} [h]
	\SetKwInOut{Input}{Input}
	\SetKwInOut{Output}{Output}
	\SetKwInOut{Initialization}{Initialization}
	\SetKwInOut{Iteration}{Iteration}
	\Input{$\mathbf{H}, \mathbf{S}, \mathbf{Y}, 	\mathbf{P}$}
	\Initialization {$q=0$, $\mathbf{R}^{0}=\mathbf{Y}$}
	\Iteration {$q \longleftarrow q+1$}	
	\SetNoFillComment
	\tcc{Activity detection}	
	${I_q}\longleftarrow \argmaxA\limits_{k, 1\leq k \leq K} \left\|\sum\limits_{l=1}^{L_d} \max \limits_{m,1 \le m \le M} \langle \diag(\mathbf{h}_{_k}) \mathbf{s}_{_{k}}^{(  \Gamma^{-1} ( \mathbf{p}_{_{k,m}} )  )}, \mathbf{r}_{_l}^{q-1} \rangle\right\|$ \\ 
	\vspace{0.1cm}
	\tcc{Data detection}
	\For{l=1 to $L_d$}
	{
		$\widehat{\mathbf{D}}_{_{I_q,l}} =  \argminA\limits_{1 \le m \le M} {\left\| {\diag(\mathbf{h}_{I_q})}\mathbf{s}_{_{I_q}}^{( \Gamma^{-1} ( \mathbf{p}_{_{I_q,m}} ) )} - \mathbf{r}_{_{l}}^{q-1} \right\|}$                              \\
		\vspace{0.1cm}
		$p=   \Gamma^{-1} (\mathbf{p}_{I_q, \widehat{\mathbf{D}}_{_{I_q,l}}}) $ \\
		\vspace{0.1cm}
		$\mathbf{r}_{_l}^{^{q}}=\mathbf{r}_{_l}^{^{q-1}}-  \diag(\mathbf{h}_{I_q})\mathbf{s}_{_{I_q}}^{(p)}$\\
	}
	\uIf{$q= K_{_{a}}$ or $\left \Vert \mathbf{r}\right \Vert < \gamma$}
	{
		stop
		\;	 
		\tcc{Decimal to binary conversion}
		$\widehat{\mathbf{B}} \longleftarrow \de (\widehat{\mathbf{D}})$
	}
	\Output{$\widehat{\mathbf{B}}$ }  
	\caption{	\label{alg:the_alg} Modified Group Matching Pursuit}
\end{algorithm}\par 
 Algorithm  \ref{alg:the_alg} takes $\mathbf{H}$, $\mathbf{S}$, $\mathbf{Y}$ and the shifting pattern matrix  $\mathbf{P} \in \mathbb{Z}^{K \times M }$ as inputs and gives the estimated  matrix $\widehat{\mathbf{B}} \in \mathbb{B}^{K\times L_c}$ as output. Each row of the matrix $\widehat{\mathbf{B}}$ represents the estimated channel coded frame of a user which consists of $L_c$ bits.  The iteration index $q$ is initialized to 0 and the residual matrix $\mathbf{R}$ is initialized to $\mathbf{Y}$.  In line 1 of the Algorithm \ref{alg:the_alg}, ${I_q}$ represents the detected active user at the   $q$-th iteration,  $\mathbf{r}_l^{q-1}\in \mathbf{R}^{q-1}$ is the $l$-th OFDM symbol of the residual at iteration $q-1$. The vector $\mathbf{s}_{_{k}}^{(  \Gamma^{-1} ( \mathbf{p}_{_{k,m}} )  )}$ is the spreading sequence of the $k$-th user which is shifted by $\Gamma^{-1} ( \mathbf{p}_{_{k,m}})$, i.e., the $m$-th entry of the shifting pattern of user $k$.  The $l$-th data symbol of user $I_q$, $\widehat{\mathbf{D}}_{_{I_q,l}} $,  is estimated in line 3 as the index of the spreading sequence in the codebook of user $I_q$ which is of  minimum Euclidean distance to the residual. In line 5, the residual is updated by subtracting the detected sequence. The algorithm terminates if the number of iterations reaches the number of active users, $K_a$, or the energy of the residual becomes less than a threshold, $\gamma$. The estimated channel coded data matrix $\widehat{\mathbf{D}}$ is then mapped to the binary  matrix $\widehat{\mathbf{B}}$.\par 
The data detection of the conventional group orthogonal matching pursuit (GOMP) algorithm is based on  estimating the transmitted sequence from the received multiuser OFDM symbols. This step involves the complex matrix pseudoinverse operation. In the modified GMP algorithm, the data detection is based on  obtaining the index of the minimum distant codeword from the received signal. Therefore, the matrix  pseudoinverse, which has complexity of the order $\mathcal{O}(NK_a^2)$ \cite{holmes2007fast}, is eliminated in the modified GMP algorithm. Furthermore, the PSK demodulation is also eliminated as the data is directly detected. However, in the activity detection the number of multiplications are increased by factor of $M$.  A table of comparison of operations required for activity and data detection in direct spreading and conventional scheme is summarized in Table \ref{tbb}.  The number of operations is obtained by calculating the number of multiplications and additions/subtractions at each step of the algorithm. For example, for a single iteration of conventional GOMP, to find the correlations between the spreading sequences and the received signal $\mathbf{Y}$, the required multiplications are $L_dNK$ as $\mathbf{S}^T\in{\mathbb{C}^{K\times N}}  $   multiples with $\mathbf{Y}\in{\mathbb{C}^{N \times L_d} } $. The number of multiplications in the first step of GOMP for detecting $K_a$ users  is therefore $L_dK_aNK$. For the modified GMP Algorithm, the number of multiplications required for detecting $K_a$ users is  $ML_dK_aNK$. Similarly after finding the number of multiplications in the residual updating step, the total number of multiplications for GOMP becomes $ L_dK_{a}N(2K+K_{a})$.

\begin{table}
	\centering
	\caption{ 	\label{tbb} Computational Complexity in the Receiver}
	\begin{tabular}{|l|l|l|}
		\hline 	
		Operations                   & Direct spreading   & Conventional scheme\\ \hline
		Multiplications              &  $ML_dK_aNK$       & $ L_dK_{a}N(2K+K_{a})$  \hspace{-0.28cm}       \\ \hline
		Additions                    & $ {K_{a}L_d(K+N+MN)}$ \hspace{-0.28cm} & $K_aL_d(K+N)$ \hspace{-0.28cm} \\ \hline
	\makecell{ \hspace{-1.2cm}Matrix \vspace{-0.1cm} \\  \hspace{-0.5cm} Pseudoinverse} \hspace{-0.33cm} & $0$          & $K_{a}$          \\ \hline
		PSK Modulation   \hspace{-0.3cm}   & No         & Yes        \\ \hline
	\end{tabular}
\end{table}\par 

\section{Performance Analysis} \label{pa}
For the performance analysis of the proposed scheme, we consider a non-orthogonal MC-CDMA uplink system.  It is assumed that each active node transmits, $L= 100$ bits per data frame. The overloading factor is defined as $\lambda=K/N$. An exponentially decaying channel with a path loss constant of two is assumed with block fading where the channel response remains same for 10 consecutive OFDM symbols.  
Other simulation parameters are summarized in Table \ref{tb}. 
\begin{table}[h]
	\centering
	\caption{Simulation parameters}
	\label{tb}
	\begin{tabular}{|l|l|}
		\hline
		Number of Nodes              & $K=1390$                \\ \hline
		Length of spreading sequence & $N=139$                \\ \hline
		Activity probability         & $p_{_a}= 0.01, 0.015$              \\ \hline
		Overloading factor           & $\lambda=10$      \\ \hline
		Channel coding               & Rate-1/2 Convolutional \\ \hline
		Modulation                   & QPSK, 8-PSK        \\ \hline
		Interleaver                  & Random              \\ \hline
		Delay spread length          & 1000 m               \\ \hline
		Fading model                 & Block fading        \\ \hline
	\end{tabular}
\end{table} \par
\begin{figure}[h]
  	\centering
  	\includegraphics[scale=.62]{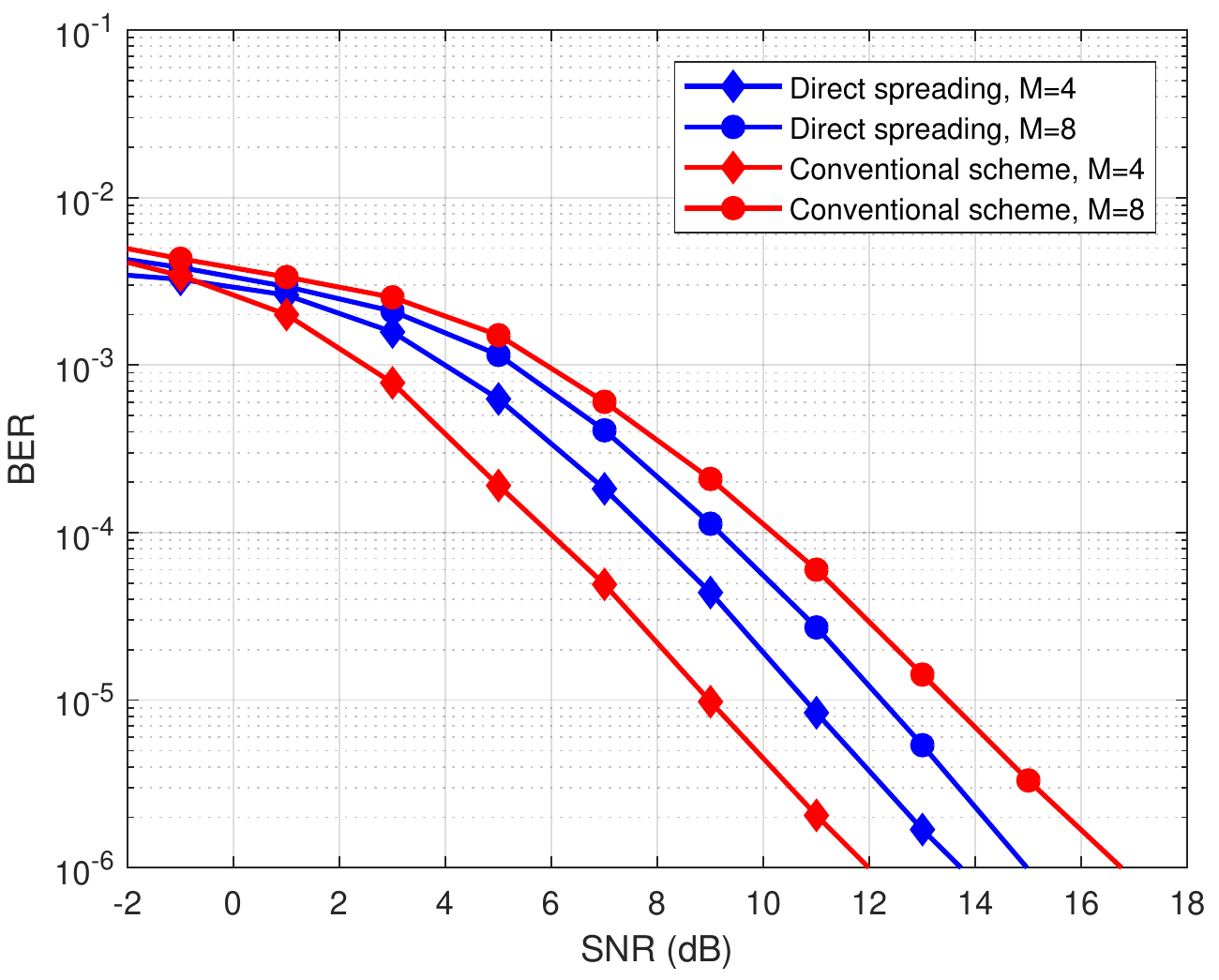}
  	\caption{Effect of increasing modulation order on conventional and direct spreading schemes: $N=139$, $\lambda= 10, \:\: p_a=0.01$ }
  	\label{exp1}
  \end{figure}

 
Figure \ref{exp1} shows the gain in the spectrum efficiency of the direct spreading scheme over the conventional scheme at modulation order of eight.  For conventional scheme, increasing the modulation order from $M = 4$ to $M = 8$, the BER significantly increases. At SNR=10 dB, it can be observed that the BER of the conventional scheme is increased by magnitude of more than 1, whereas the increase in BER of the proposed scheme is less  than half a magnitude. At lower SNR, especially below 0 dB, the BER performance is nearly same for both the schemes at $M=8$. The reason is that, for SNR $>$ 10 dB, the minimum Euclidean distance, $d_{min}$, is the main factor that determines the performance. As discussed in Section \ref {proposed},  increasing modulation order from $M=4$ to 8,  the impact on  $d_{min}$ of the multidimensional codebook is less than that of  the PSK constellation, therefore,  a gain of more than 1 dB is achieved over the conventional scheme at  BER = $10^{-5}$. For $M = 4$, $d_{min}$ of the PSK modulator is greater than that of the multidimensional codebooks, which results in the performance gain of the conventional scheme. The gain of the proposed scheme at higher modulation schemes therefore validates the theoretical justification in Section \ref{proposed}.\par
\begin{figure}[!h]
	\centering
	\includegraphics[scale=.62]{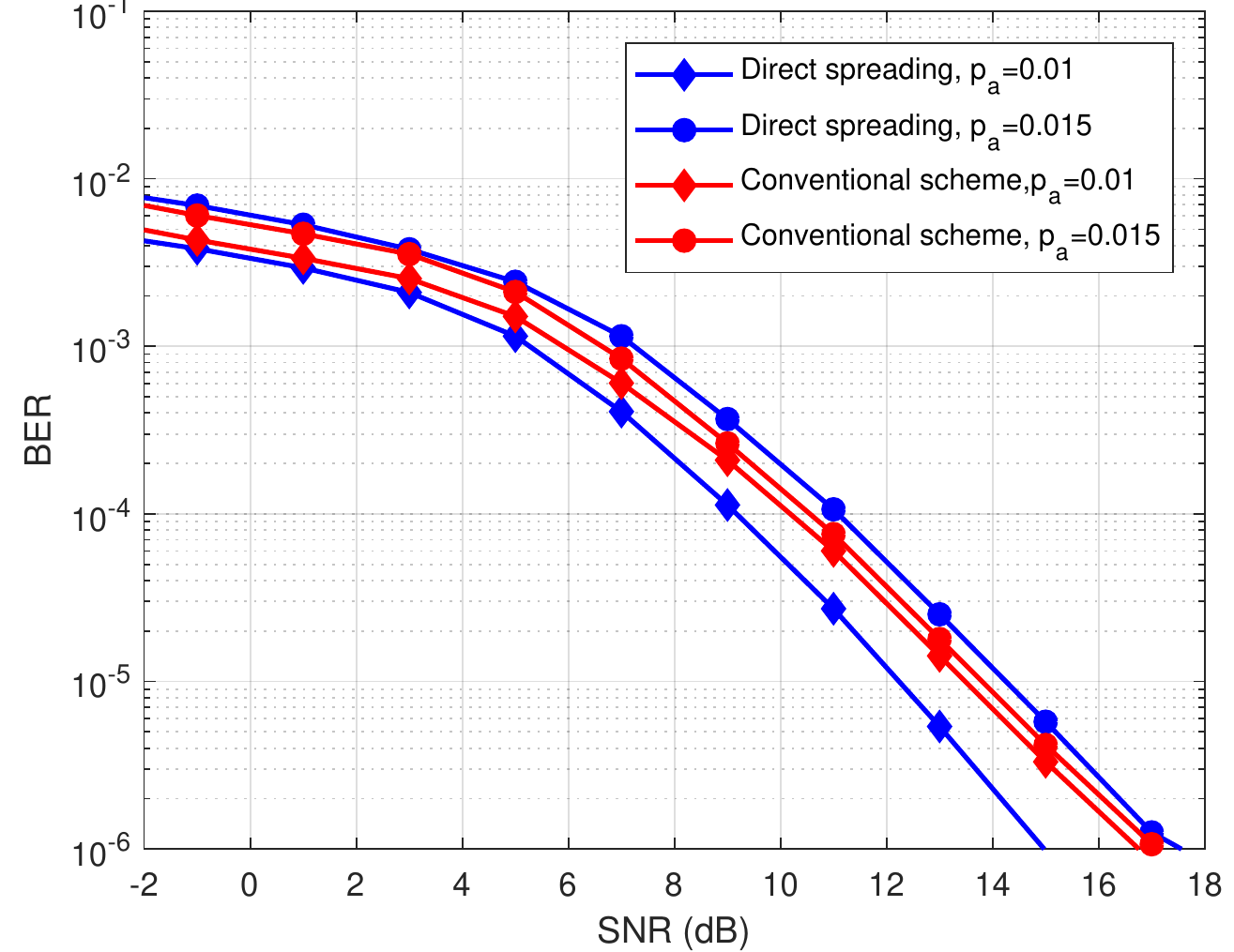}
		\caption{Effect of increasing activity probability on conventional and direct spreading schemes: $N=139$, $\lambda= 10, \:\: M=8$}
	\label{exp3}
\end{figure}

Figure \ref{exp3} presents the effect of increasing the activity probability on the performance of the direct and conventional spreading schemes. It can be observed that at activity probability of 1.5\%, the performance of both schemes is nearly same. The results in Figure \ref{exp3}  shows  that using the proposed scheme, we can achieve the same performance as that of the PSK based scheme even at higher activity with a simple and low power processing at the transmitting node. For increasing the activity probability from 1\% to 1.5\%, it is also observed that the direct spreading scheme is more prone to errors when large number of nodes is active. However, for mMTC the activity probability is considerably low, therefore, the direct spreading scheme is practical and can be used for mMTC.  
\section{Conclusion and future work} \label {con} 
In this paper, a codebook based non-orthogonal MC-CDMA is proposed, which combines the PSK modulator and the spreader to a single symbol-to-sequence spreader. The symbol-to-sequence spreader directly maps the input bits into a multidimensional codeword in the user specific codebook. Compressive sensing based multiuser detection is used at the receiver to detect the activity and the corresponding data. The proposed scheme reduces the performance loss for increasing the modulation order and therefore outperforms the conventional scheme at higher modulation order. The performance gain comes from the fact that for increasing the modulation order the reduction in the minimum Euclidean distance of the multidimensional codebook is less than that of the PSK modulator.  Moreover, eliminating the PSK modulator simplifies the data processing at the sensor node, which contributes in lowering the  power consumption, consequently, increasing the battery life of the sensor node.­\par
The current results are based on the randomly generated base spreading sequences; therefore, designing deterministic codebooks, which have minimum inter-codebook as well as intra-codebook correlation, will be an interesting topic to investigate. Furthermore, designing more efficient decoding algorithms for non-orthogonal codebook based MC-CDMA in dense scenario is also an interesting area to explore.
 \bibliographystyle{unsrt}
\bibliography{bibfile} 
  \end{document}